\documentclass[aps,prl,twocolumn,floatfix,superscriptaddress,reprint,longbibliography]{revtex4-1}
\usepackage[utf8]{inputenc}
\usepackage{graphicx}
\usepackage{amsmath}
\usepackage{siunitx}
\usepackage[usenames,dvipsnames]{xcolor}
\usepackage{tikz}  % used for circled
\usepackage{commath}
\usepackage{ulem}
\definecolor{MyBlue}{rgb}{0,0.3,0.6}
\usepackage[colorlinks=true,linkcolor=MyBlue,plainpages=false,citecolor=MyBlue,urlcolor=MyBlue]{hyperref}

\definecolor{mygreen}{rgb}{0,0.65,0}
\definecolor{mygreen2}{rgb}{1,0,0}

\newcommand{\makered}[1]{\textcolor{black}{#1}}

\DeclareRobustCommand\circled[1]{%
            \tikz[
              baseline=-1mm,
            ]{%
                \clip (0mm,-0.05mm) circle [radius=1.55mm];
                \draw[black, line width=0.2mm]
                   (0mm,-0.05mm) circle [radius=1.5mm];%
                \node[black, anchor=center, scale=0.75, yshift=0.05mm]
                (0mm,0mm) {#1};%
            }}%
\newcommand{\Ta}{\text{Ta}}

\newcommand{\Nuw}{\text{Nu}_\omega}
\newcommand{\nue}{\nu_\text{eff}}
\newcommand{\omegai}{\omega_\text{ic}}
\newcommand{\omegao}{\omega_\text{oc}}
\newcommand{\ri}{r_\text{ic}}
\newcommand{\ro}{r_\text{oc}}

\newcommand{\reff}[1]{Fig.~\ref{#1}}

\begin{document}
\title{Catastrophic phase inversion in high-Reynolds number turbulent Taylor--Couette flow}

\author{Dennis~Bakhuis}
\affiliation{Physics of Fluids Group, Max Planck UT Center for Complex Fluid Dynamics,\break
MESA+ Institute and J.M. Burgers Centre for Fluid Dynamics, University of Twente, 7500 AE Enschede, The Netherlands}
\author{Rodrigo~Ezeta}
\affiliation{Physics of Fluids Group, Max Planck UT Center for Complex Fluid Dynamics,\break
MESA+ Institute and J.M. Burgers Centre for Fluid Dynamics, University of Twente, 7500 AE Enschede, The Netherlands}

\author{Pim~A.~Bullee}
\affiliation{Physics of Fluids Group, Max Planck UT Center for Complex Fluid Dynamics,\break
MESA+ Institute and J.M. Burgers Centre for Fluid Dynamics, University of Twente, 7500 AE Enschede, The Netherlands}
\affiliation{Soft matter, Fluidics and Interfaces, MESA+ Institute for
Nanotechnology, University of Twente, 7500 AE Enschede, The Netherlands}

\author{Alvaro~Marin}
\affiliation{Physics of Fluids Group, Max Planck UT Center for Complex Fluid Dynamics,\break
MESA+ Institute and J.M. Burgers Centre for Fluid Dynamics, University of Twente, 7500 AE Enschede, The Netherlands}

\author{Detlef~Lohse}
\email{d.lohse@utwente.nl}
\affiliation{Physics of Fluids Group, Max Planck UT Center for Complex Fluid Dynamics,\break
MESA+ Institute and J.M. Burgers Centre for Fluid Dynamics, University of Twente, 7500 AE Enschede, The Netherlands}
\affiliation{Max Planck Institute for Dynamics and Self-Organization, Am Fa\ss berg 17, 37077 G\"{o}ttingen, Germany}

\author{Chao~Sun}
\email{chaosun@tsinghua.edu.cn}
\affiliation{Center for Combustion Energy, Key Laboratory for Thermal Science
and Power Engineering of Ministry of Education, Department of Energy and Power
Engineering, Tsinghua University, Beijing 100084, China}
\affiliation{Department of Engineering Mechanics, School of Aerospace Engineering, Tsinghua University, Beijing 100084, China}

\author{Sander~G.~Huisman}
\email{s.g.huisman@gmail.com}
\affiliation{Physics of Fluids Group, Max Planck UT Center for Complex Fluid Dynamics,\break
MESA+ Institute and J.M. Burgers Centre for Fluid Dynamics, University of Twente, 7500 AE Enschede, The Netherlands}

\date{\today}

\begin{abstract}
\noindent Emulsions are omnipresent in the food industry, health care, and chemical synthesis. In this Letter the dynamics of meta-stable oil-water emulsions in highly turbulent ($10^{11}\leq\text{Ta}\leq 3\times 10^{13}$) Taylor--Couette flow, far from equilibrium, is investigated. By varying the oil-in-water void fraction, catastrophic phase inversion between oil-in-water and water-in-oil emulsions can be triggered, changing the morphology, including droplet sizes, and rheological properties of the mixture, dramatically. The manifestation of these different states is exemplified by combining global torque measurements and local in-situ laser induced fluorescence (LIF) microscopy imaging. Despite the turbulent state of the flow and the dynamic equilibrium of the oil-water mixture, the global torque response of the system is found to be as if the fluid were Newtonian, and the effective viscosity of the mixture was found to be several times bigger or smaller than either of its constituents.
\end{abstract}
\maketitle

\noindent Mixtures of oil and water are omnipresent in petrochemical processes \cite{Petrochemical2001}, pharmaceutics \cite{Sarker2013}, as well as in the food industry \cite{mcClements2004}. For example, in oil recovery, water is generally used as a carrier liquid to extract oil from the ground, creating an emulsion \cite{petro2015}. Also in the further processing of the crude oil, and the synthesis of other chemical compounds, emulsions are a common occurrence. As water and oil are immiscible, due to their polar and non-polar nature, respectively, they fully separate in two phases. However, when vigorously stirred by turbulence, they can be dispersed into each other. One phase is fragmented to form drops (becoming the dispersed phase) and suspended inside the other liquid (continuous phase), creating an emulsion. Simple emulsions come in two forms: oil droplets suspended in water (o--w) or water droplets suspended in oil (w--o). Without continuous stirring, however, both types of emulsions are unstable as density differences and buoyancy promote coalescence of the dispersed phase, causing the two phases to separate. Depending on the application, emulsifiers (stabilisers, surfactants, polymers, or an amphiphile) can be added to gain stability against coalescence---frequently done in the food industry using \textit{e.g.}~egg yolk or its active component lecithin. Mayonnaise may be the most famous example of such a stabilized emulsion.

In various industrial processes, emulsions are pumped through pipes \cite{plasencia2013} and stirred in tanks, and determining the rheological properties of these emulsions in these turbulent flow conditions is hard. Conventional rheometers work in the laminar regime such that their flow profile can be easily derived and used for calibration. However, emulsions, without stabilisers, would quickly phase-separate in such rheometers, therefore not accurately replicating the meta-stable state---a state far from equilibrium---that the emulsion has in a turbulent flow. In such turbulent flow of a meta-stable emulsion, the dispersed phase is continuously broken up by the eddies while at the same time drops continuously coalesce, creating a dynamic equilibrium. Stopping the energy input quickly destroys the dynamic equilibrium and the phases separate.

The theory of the break-up of droplets in a turbulent flow was pioneered by Kolmogorov \cite{Kolmogorov1949} and Hinze \cite{Hinze1955}. Their focus was to determine a correlation between the flow scales and the average droplet diameter by dimensional analysis. Turbulent eddies with a scale similar to the droplet can destabilize the interfaces, leading to the break-up of droplets \cite{Andersson2006}. While the smallest scale in a turbulent flow, \textit{i.e.}~the Kolmogorov scale $\eta_K$, was hypothesized as the lower bound for the droplet diameter, it was experimentally found that a portion of droplets can have a smaller size \cite{Zhou1998}. At these scales (viscous subrange), sub-eddy viscous stresses dominate over inertial stresses, making smaller droplets possible \cite{Shinnar1961,Boxall2011}.

The stability of an emulsion is not only dictated by the presence of a suitable emulsifier, but also the relative volume fractions are important. For increasing oil fractions, it becomes harder and harder to maintain an oil-in-water emulsion (without using an emulsifier), and finally oil and water switch their roles and the system undergoes a so-called phase-inversion \cite{Salager2000, Perazzo2015} to the water droplet in oil case. Various attempts at modeling the inversion point of this phase-inversion have been undertaken: minimal dissipation models \cite{Poesio2008, Ngan2009}, energy barrier models \cite{Piela2009}, or coalescence/breakup models \cite{Brauner2002,Yeo2002}. The critical inversion point of o--w to w--o and w--o to o--w are generally not at the same oil volume fraction, i.e.~hysteresis is observed \cite{Brauner2002,Yeo2002,Piela2006,Moradpour2011}. According to an extended Ginzburg-Landau model \cite{Piela2009}, the width of the hysteresis region, also known as the ambivalence region, is a unique property of the emulsion mixture and independent of t\makered{h}e Reynolds, Froude, and Weber numbers of the system. Lastly, the typical time-scale of the inversion can vary between several days, or being nearly instantaneous; in the latter case the phase inversion is called a \textit{catastrophic} phase inversion \cite{Dickinson1981,Vaessen1996,Tyrode2005}.

In this Letter we investigate the dynamic phase stability and the rheological properties of an unstable emulsion without emulsifiers in highly turbulent shear flow. To do so, we make use of recent advances \cite{vanGils2011,Grossmann2016} in large scale turbulent Taylor--Couette (TC) facilities, which make it possible to use this geometry as a rheometer---classically solely used in the laminar regime---even in the turbulent regime.

\begin{figure}
\centering
\includegraphics[width=\columnwidth]{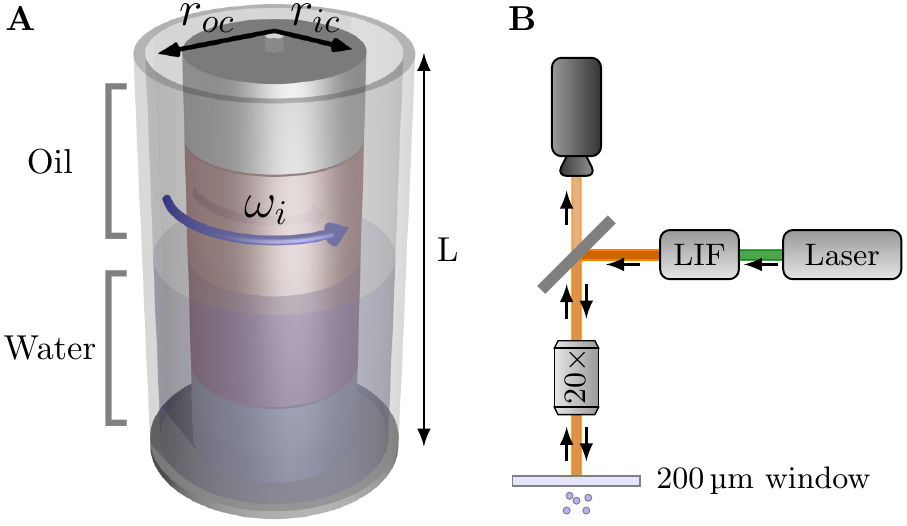}
\caption{A. Schematic of the T$^3$C setup. Initially, the water (bottom) and oil (top) are separated. B. In-situ high-speed microscopy using LIF lighting, a half-mirror, and a $20\times$ magnification lens, through a \SI{200}{\micro\meter} thick window at the top of the setup shown in A.}
\label{fig:setupoil}
\end{figure}

We employ the Twente Turbulent Taylor--Couette facility \cite{vanGils2011}, as it provides local and global measurement access in a controlled and closed geometry, see \reff{fig:setupoil}A.
The apparatus has an inner cylinder (IC) radius $\ri=\SI{200}{\milli\meter}$, an outer cylinder (OC) radius $\ro=\SI{279.4}{\milli\meter}$, and a height $L=\SI{927}{\milli\meter}$, resulting in a gap width $\delta=\SI{79.4}{\milli\meter}$, a radius ratio $\eta=\ri/\ro=0.716$, and an aspect ratio $\Gamma=L/\delta = 11.7$. The torque, $\mathcal{T}$, required to rotate the IC at constant angular velocity $\omegai$ is measured using a torque sensor. The temperature of the fluids is kept within \SI[separate-uncertainty = true,multi-part-units=single]{21.0(5)}{\celsius}. Optical access to the flow is through a window on top of the system, modified to hold a \SI{200}{\micro\metre} thick microscope slide, thereby allowing to operate a microscopy system (shown in \reff{fig:setupoil}B). The microscopy system consists of LIF lighting, a half-mirror, $20\times$ magnification lens, and a camera \cite{camera}. We make use of demineralised water and a low-viscosity silicone oil \cite{ShinEtsuKF96} with $\nu_o=\SI{1.03}{\milli\meter\squared\per\second}$ at $\SI{25}{\celsius}$ and an interfacial tension with water of $\gamma =\SI{42.7}{\milli\newton\per\metre}$ \cite{Kanellopoulos1971}.\\

The driving strength of TC flow can be quantified by the Taylor number \cite{Eckhardt2007}:
\begin{align}
    \text{Ta} = \frac 14 \sigma \delta^2 (\ri + \ro)^2 (\omegai - \omegao)^2 /
    \nu^2
    \label{eq:oilTa}
\end{align}
where $\sigma = ((1+\eta)/(2\sqrt \eta))^4$ is a geometric constant. The response of the system, the torque, can then be captured as an angular velocity Nusselt number \cite{Eckhardt2007}:
\begin{align}
    \text{Nu}_\omega \equiv  \frac{J_\omega}{J^{\text{lam}}_{\omega}} =
    \frac{\mathcal{T}}{2\pi L \rho J^{\text{lam}}_{\omega}}.
    \label{eq:oilNu}
\end{align}
Here $J^{\text{lam}}_{\omega} = 2 \nu \ri^2 \ro^2(\omegai - \omegao)/(\ro^2 - \ri^2)$ is the angular velocity transport for laminar, non-vortical flow. From dimensional analysis $\text{Nu}_\omega=f(\text{Ta},\omegao/\omegai,\Gamma,\eta)$, where $f$ is some general function, which has been experimentally found and numerically verified \cite{vanGils2011b,Paoletti2011,Huisman2012b, Ostilla-Monico2014b, Huisman2014,Grossmann2016}. We keep $\eta$ and $\Gamma$ constant and the OC remains stationary, $\omegao=0$, such that we have $\text{Nu}_\omega=f(\text{Ta})$.

In order to compute an effective viscosity for a meta-stable emulsions (MSE), we will assume that the system globally responds as if it were filled with a Newtonian liquid at these shear rates. While quasi-statically ramping up the inner cylinder from $\omegai/(2\pi)=\SI{4}{\hertz}$ to $\SI{20}{\hertz}$, the torque $\mathcal{T}$ was measured for various oil volume fractions $\alpha=V_o/(V_o+V_w)$, where $V_o$ ($V_w$) is the volume of the oil (water) phase. Experiments based on liquids with known viscosity define the curve $\Nuw=f(\Ta)$ for our geometry, which we do so using the $\alpha=0\%$ and $\alpha=100\%$ cases. The scaling of $\Nuw=f(\Ta)$ can now be exploited to determine $\nue(\alpha)$ such as to collapse all the data on to a single curve.
\begin{figure}[t]
\centering
\includegraphics{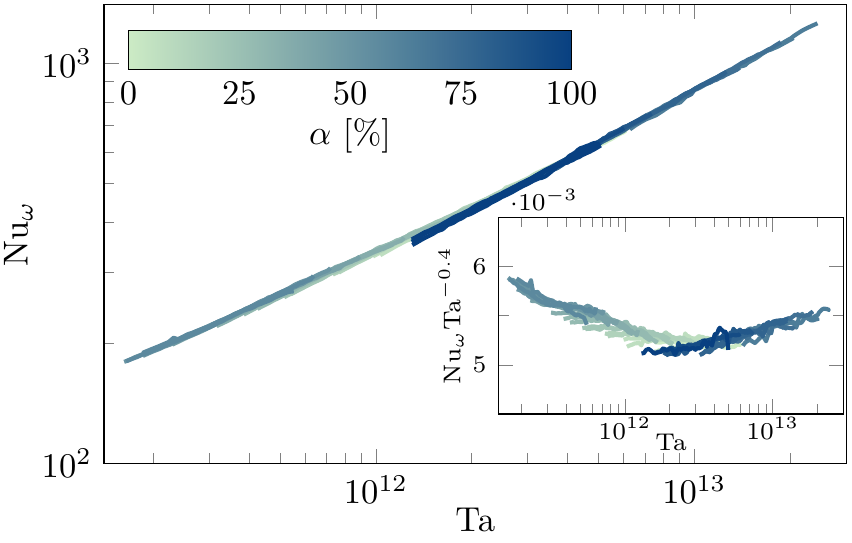}
\caption{
    $\text{Nu}_\omega$ as a function of $\text{Ta}$ for a total of 32 experiments. By exploiting the relation between $\Nuw=f(\Ta)$, an effective viscosity $\nue$ was found for each experiment such that all the datasets collapse on a single curve, by minimizing $\sigma(\Nuw)$ for the collective data, binned in $\Ta$. The inset shows the compensated datasets, revealing the overlap within a couple of percent. The colors of the lines indicate the oil fraction.
}
\label{fig:torqueoil}
\end{figure}
Indeed, all 32 datasets collapse onto a single universal curve (see~\reff{fig:torqueoil}), consistent with previous studies on Newtonian liquids and emulsions \cite{wen33,Lathrop1992,lew99,ravelet2007,ravelet2008,Paoletti2011,vanGils2011b,Huisman2014, Ostilla-Monico2014b}. To appreciate the quality of the collapse, all data were compensated by $\Ta^{0.4}$, which is the effective scaling around $\Ta=\mathcal{O}(10^{12})$ (inset of \reff{fig:torqueoil}). This analysis shows that all datasets combined have a standard deviation of only $1.2\%$, and that the expected scaling is also around $\Nuw \propto \Ta^{0.4}$, which justifies the Newtonian assumption made earlier for the MSE at such large shear rates.

\begin{figure}[t]
    \centering
    \includegraphics{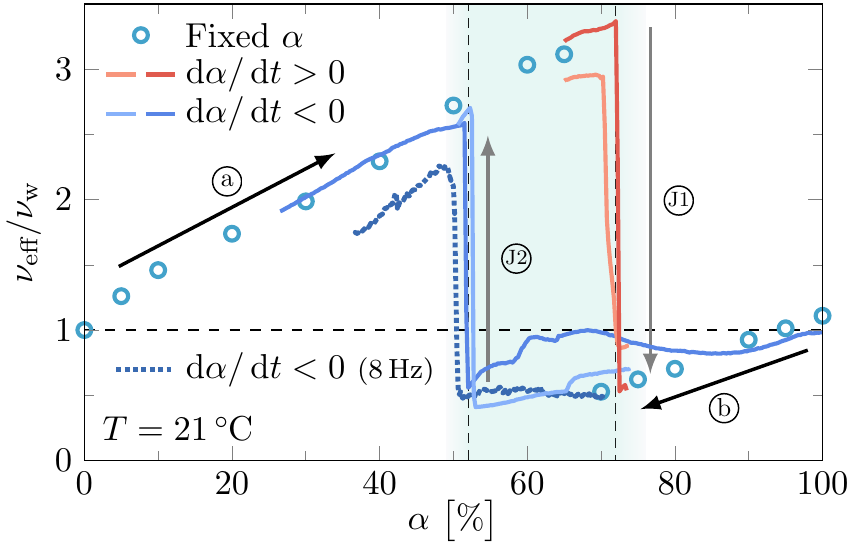}
    \caption{Effective viscosity normalized by the viscosity of water as function of oil volume fraction. The measurements with fixed $\alpha$ are shown by the circles. The continuous measurements, for which $\alpha$ is changed during the experiments ($\omega_{ic}/(2\pi)$ is fixed at \SI{17.5}{\hertz}), are shown as solid lines, while the dotted line is an experiment with $\omega_{ic}/(2\pi)=\SI{8}{\hertz}$.
    The shaded area between the dashed vertical lines shows the ambivalence region, bounded by two catastrophic phase inversion, \circled{\footnotesize{J1}}: w--o $\rightarrow$ o--w and \circled{\footnotesize{J2}}: o--w $\rightarrow$ w--o. When water is the
    continuous phase, increasing $\alpha$ (\circled{a}) leads to an increase in $\nue$, while when oil is the continuous phase decreasing $\alpha$ (\circled{b}), decreases $\nue$.}
    \label{fig:viscosity}
\end{figure}

The effective viscosity $\nue(\alpha)$, normalized using the viscosity of water $\nu_w$, is shown in \reff{fig:viscosity} as blue circles. Each experimental measurement in this set is performed at a constant oil fraction $\alpha$, while ensuring that the two phases were fully mixed, and is depicted as `fixed $\alpha$' in the figure. We observe two disconnected branches, the left branch for $\alpha \leq 65\%$, and the right branch for $\alpha\geq 70\%$, corresponding to o--w and w--o emulsions, respectively.
Remarkably, we find that by adding oil (\circled{a}) the effective viscosity increases \textit{beyond} the viscosities of each of the constituents, making the fluid three times as viscous as water for $\alpha=65\%$. However, further increasing $\alpha$ in excess of $65\%$ results in a dramatic drop in $\nue$, caused by the catastrophic phase inversion. Around the phase inversion $65\%<\alpha<70\%$, $\nue$ decreases by a factor of 6, resulting in a tremendous change of more than $40\%$ in torque. Also, compared to the case of pure oil ($\alpha=100\%$), we find that the case of $\alpha=70\%$ has a lower effective viscosity, similar to other w--o emulsions \cite{Pal1993}. We further note that our emulsions do not adhere to Einstein's \cite{einstein1906} or Taylor's \cite{taylor1932} viscosity correction factors around $\alpha=0\%$ (where the viscosity increases faster than predicted) or $\alpha=100\%$ (where the viscosity decreases rather than increases).

To further investigate what happens at the phase inversion we now quasi-statically increase and decrease $\alpha$, while keeping $\omegai/(2\pi)$ fixed to \SI{17.5}{\hertz}, by slowly draining the setup at the bottom while filling it with oil or water from the top, see the red and blue curves in \reff{fig:viscosity}. The instantaneous value of $\alpha$ is obtained by recording the mass of the injected liquid and assuming that the liquid drained is a homogeneously mixed emulsion. At the end of the experiment, $\alpha$ was directly measured and found to be within $1\%$ of the calculated value. We find that for both $\dif{\alpha}/\dif{t} > 0$ (red curves) and $\dif{\alpha}/\dif{t} < 0$ (blue curves) we see a catastrophic phase-inversion for $\alpha=72\%$ (\circled{\footnotesize{J1}}) and $\alpha=52\%$ (\circled{\footnotesize{J2}}), respectively. The measurements for fixed $\alpha$ agree with those where we continuously change $\alpha$, though the high $\nue$ branch (o--w) was chosen by the system inside the ambivalence region. The emergence of the ambivalence region $52\% \leq \alpha \leq 72\%$ (shaded blue area in between vertical dashed lines in \reff{fig:viscosity}), is unexpected to exist for such high $\text{Ta}$ (or equivalent Reynolds numbers of $\mathcal{O}(10^6)$). One would expect that for such high Taylor numbers the system provides ample kinetic energy to freely explore the phase-space, and thus arrive at the energetically most favorable state; however it has been shown before that turbulent TC flow is susceptible to hysteresis \cite{and86,Huisman2014,vanderVeen2016,gul2018} and that emulsions also show hysteretic behavior \cite{sjoblom2005}.
For turbulent pipe flows \cite{Pal1993,Piela2006,Piela2008} the width of the ambivalence region solely depends on the ratio of the dispersed phase injection rate and the total flow rate. We find that, for an injection rate of the dispersed phase of approximately \SI{12.5}{\milli\liter\per\second}, the ambivalence region gets slightly wider when $\Ta$ is decreased (illustrated from dashed to dotted boundaries in \reff{fig:viscosity}). This widening of the ambivalence region is thus delaying the phase inversion for both o--w and w--o, in contrast to the prediction of the extended Ginzburg-Landau model \cite{Piela2009}, from which one would expect independence of the Taylor number.

Which branch in the ambivalence region is chosen depends on the trajectory in parameter space used to reach $(\alpha,\omegai)$. Repeated experiment are found to give us repeatable results, such that a certain emulsion state (w--o or o--w) (and its corresponding low or high $\nue$) can be selected by performing the corresponding trajectory in phase-space, opening possibilities for drag reduction for the transport of emulsions that are within the ambivalence region. We label these phase-inversions as catastrophic as these inversions completely change the morphology of the flow within a second, which results in a sharp increase (or decrease) of the torque by $\approx 35\%$. For both phase inversion \circled{\footnotesize{J1}} and \circled{\footnotesize{J2}} we see a sharp jump with the system, switching from a drag reduction to a severe drag increase, or vice versa.

The different responses seen in each of the branches in Figure \ref{fig:viscosity} demand a local view of the flow. Adding inclusions such as particles, bubbles, or immiscible fluids to a liquid can dramatically change its rheology \cite{Derkach2009, Stickel2005}. Solids generally increase the effective viscosity \cite{einstein1906,Bakhuis2018}, while small amounts of polymers \cite{White2008}, oil without surfactant \cite{Pal1993}, or gas \cite{vandenBerg2007,vanGils2013,Verschoof2016} are known to reduce drag, yielding a lower effective viscosity than the original liquid phase. For the case of air-lubrication, the current understanding suggests \cite{lu2005,Verschoof2016,vanGils2013,Spandan2018} that the Weber number, describing the deformability of the bubble, is of importance. The decrease in $\nue$ for w--o emulsions (\circled{b}) could be a similar process as in bubble drag reduction \cite{vandenBerg2007,vanGils2013,Verschoof2016}, related to the deformability of the dispersed phase, for which large droplets are required. It was already found before that the droplet size in emulsions have a large impact on the rheology \cite{Pal1996}. And, using the same reasoning, an increasing effective viscosity $\nue$ (\circled{a}) could be connected with the presence of small and non-deformable droplets which could act as solid-like particles and therefore increase $\nue$ \cite{Stickel2005,Bakhuis2018,devita2019}.

Analyzing the size of our droplets is therefore paramount in order to see whether or not the above-mentioned effects could be of importance. Due to the meta-stable nature of the emulsion it quickly separates when driving is not continuously supplied, and therefore samples taken from the system and analyzed under a standard microscope do not have the same droplet sizes as those in the system while it is in operation. Consequently, we choose to bring the microscope to our apparatus and size the dispersed phase while in operation.

\begin{figure}
\centering
\includegraphics{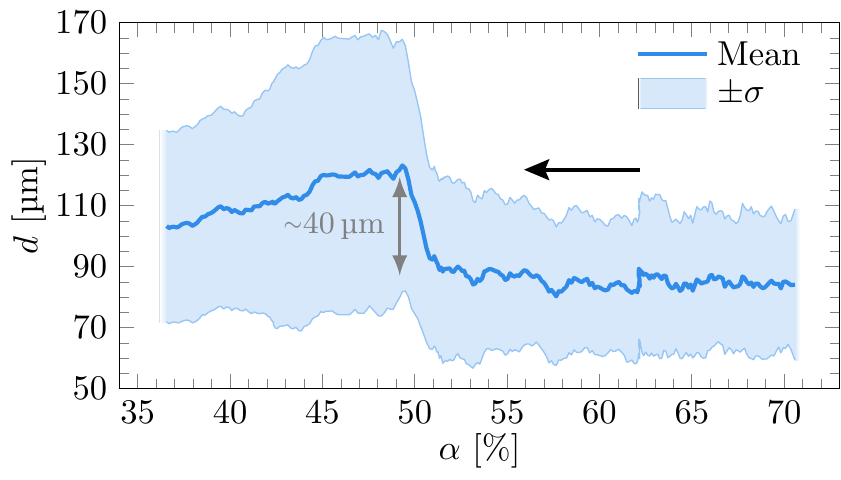}
\caption{Droplet diameter of the dispersed phase obtained using in-situ macroscopy for decreasing $\alpha$ for a rotational speed of $\omegai/(2\pi) = \SI{8}{\hertz}$. A sharp jump in average diameter is found at the inversion point $\alpha\approx 50\%$, as is also visible in the measured torque (and from that $\nue$) as dotted line in \reff{fig:viscosity}. The arrow indicates the direction of the $\alpha$-modification. Two example images are given in the supplemental material.}
\label{fig:dropletsize}
\end{figure}

The optical arrangement is shown in \reff{fig:setupoil}B. The dispersed droplets were visualized by the reflected light from their interface and no fluorescent dye was employed. We performed a sweep of $\alpha$ from $70\%$ to $37\%$ over the course of \SI{16.5}{\min} at a constant angular velocity of $\omegai/(2\pi) = \SI{8}{\hertz}$ while recording the torque and continuously imaging the droplets, see the dashed line in \reff{fig:viscosity} for the $\nue$ and \reff{fig:dropletsize} for the diameter of the droplets in the dispersed phase.
A total of 14500 frames were recorded and for each frame all the droplets were identified. We only observed simple emulsions and did not find any multiple emulsions (emulsions with deeper embeddings, e.g.~w--o--w) \cite{Jahanzad2009}. The phase inversion can be seen in the dramatic jump in $\nue$ in \reff{fig:viscosity}, but also directly in the mean diameter $d$ of the dispersed phase in \reff{fig:dropletsize}. We find that the phase inversion causes a $4\times$ jump in $\nue$ and the oil droplets were about $50\%$ larger in radius (about $3.5\times$ larger in volume) than the water droplets just before and after the phase inversion.

As in previous research \cite{vandenBerg2007,vanGils2013,Verschoof2016}, we calculate the Weber number to gauge the deformability and associated mobility of the droplets. In the turbulent case $\text{We}=\rho u'^2 d / \gamma$, where $\rho$ is the density of the mixture and $u'$ is an estimate of the azimuthal velocity fluctuations based on previous research \cite{ezeta2018}. We find that $\text{We}$ is $0.07$ just before the jump, and $0.14$ just after the jump, both well below unity. The requirement for bubbly drag reduction was found to be $\text{We}>1$ \cite{vandenBerg2007,vanGils2013,Verschoof2016}, so the traditional drag-reduction mechanism, invoking deformability of the bubbles, is likely not applicable for our high oil void fraction emulsions.

The size of the droplets is linked to the underlying flow structure, whose relevant length scale is the Kolmogorov length scale: $\eta_K=\left(\nu^3/\epsilon \right)^{1/4}$, comparing inertia and viscosity. Here $\epsilon$ is the turbulent energy dissipation rate. From this relation we find that $\eta_K=\SI{12}{\micro\meter}$ and \SI{33}{\micro\meter} just before and after the jump. These values are both considerably smaller than the droplets we encountered.
The relevant length scale in case of a dispersion in fact is the Hinze length scale~\cite{Hinze1955}, which compares inertia and capilarity: \makered{$d_{\text{Hinze}} \approx 0.725 \left(\gamma/\rho_{\text{cont}}\right)^{3/5} \epsilon^{-2/5} \approx \SI{1320}{\micro\meter}$ and $\SI{967}{\micro\meter}$ just before and after the jump, respectively.} This is larger than what we optically found for the droplets. \makered{This estimate is based on dimensional analysis and different before and after the jump as $\rho_{\text{cont}}$ of the carrier fluid abruptly changes from oil to water and $\epsilon$ also changes during the jump.} Therefore, the first challenge is to understand the reason for the asymmetric jump in drag at the phase inversion.
On the one hand, density differences might allow the heavier oil droplets to migrate closer to the IC than to the OC, while the lighter water droplets would do the opposite. On the other hand, oil adhesion to the metallic IC surface is known to increase when surrounded by water \cite{clint2001adhesion}, while water droplets are not expected to adhere to the acrylic OC surface surrounded by oil. These are two strong factors which could explain the asymmetry in the observed torque. Direct observation of these effects, to verify or falsify these hypotheses, are unfortunately impossible in the current experimental facility.

In summary, in this Letter we have studied the flow of meta-stable emulsions in an intensely sheared rotating flow. Exploiting the known scaling of the ultimate regime of Taylor--Couette flow, an effective viscosity of the emulsion can be measured, and thereby using the TC apparatus as a rheometer far beyond the conventional regime. With water as the continuous phase, the addition of oil droplets increases the drag of the system. Interestingly, drag is reduced when the water is injected in a continuous oil phase.
At a critical volume fraction, a catastrophic phase inversion takes place, which dramatically changes the rheological properties.
Using in-situ microscopy we were able to obtain droplet sizes in the TC system while in operation. This showed that the jump observed in the torque can also be seen in the droplet size of the dispersed phase. The complex morphology on a microscopic scale with a wide size distribution of the dispersed phase was found to globally behave on the torque as if it were a Newtonian fluid. Our results demonstrate that we can select the emulsion type in the ambivalence region, which opens ways to \makered{achieve} major drag reductions for the transport of emulsions. Our finding that an emulsion at high turbulent intensity behaves as if it were a Newtonian fluid is also highly relevant \makered{to} the transport and mixing of emulsions, \textit{e.g.}~in oil recovery and the petrochemical industry.

\begin{acknowledgments}
We thank Raymond Bergmann, Nicolas Bremond, Sascha Hilgenfeldt, Dominik Krug, Henri Lhuissier, Andrea Prosperetti, Rian Ruhl, Vamsi Spandan, Peter Veenstra, Ruben Verschoof, Doris Vollmer, Jelle Will, and Jeff Wood for various stimulating discussions. Raymond Kip for help in the lab for preliminary work. Gert-Wim Bruggert and Martin Bos for technical support. We also thank Guillaume Lajoinie and Tim Segers for assisting setting up the in-situ microscopy measurements. This work was supported by Natural Science Foundation of China under grant nos 11988102, 91852202, 11861131005 and 11672156, the Netherlands Organisation for Scientific Research (NWO) under VIDI Grant No. 13477, STW, FOM, and MCEC.
\end{acknowledgments}

\end{document}